\documentclass[11pt]{article}
\usepackage{amssymb}
\usepackage{graphics}
\usepackage{epsfig}
\usepackage{a4wide}
\usepackage{color}

\begin{document}

\newcommand{\ppww}{$pp \to W^+ W^-+X~$ }
\newcommand{\ppwwg}{$pp \to W^+ W^-g+X~$ }
\newcommand{\qqww}{$q\bar q \to W^+W^-~$}
\newcommand{\qqwwg}{$q\bar q \to W^+W^-g~$}
\newcommand{\ggww}{$gg \to W^+W^-~$}
\newcommand{\ggwwg}{$gg \to W^+W^-g~$}
\newcommand{\qgwwq}{$q(\bar q)g \to W^+W^-q(\bar q)~$}
\newcommand{\ppuuww}{$pp \to u\bar u \to W^+W^-+X~$}
\newcommand{\ppqqww}{$pp \to q\bar q \to W^+W^-+X~$}
\newcommand{\ppggww}{$pp \to gg \to W^+W^-+X~$}
\newcommand{\ppuuwwg}{$pp \to u\bar u \to W^+W^-g+X~$}
\newcommand{\ppwwllvv}{$pp \to W^+W^- \to l^+\nu_ll^-\bar{\nu}_l + X~$}
\newcommand{\ppwlv}{$pp \to W^+W^- \to W^{\pm}l^{\mp}\stackrel{(-)}{\nu_l} + X~$}

\title{ Revisiting the large extra dimension effects on $W$-pair production at the LHC in NLO QCD }
\author{ \textcolor{red}{}
Bai Yu-Ming, Guo Lei, Li Xiao-Zhou, Ma Wen-Gan, and Zhang Ren-You \\
{\small Department of Modern Physics, University of Science and Technology}\\
{\small of China (USTC), Hefei, Anhui 230026, People's Republic of China}  }

\date{}
\maketitle \vskip 15mm
\begin{abstract}
In the framework of the large extra dimensions (LED) model, we
investigate the effects induced by the Kaluza-Klein gravitons
up to the QCD next-to-leading order (NLO) on the $W$-pair production
followed by a subsequential $W$-decay at the CERN LHC. We depict the
regions in the $\mathcal{L}-M_{S}$ parameter space where the LED
effect can and cannot be observed from the analyses of the $pp \to
W^+W^- + X$ and $pp \to W^+W^- \to W^{\pm}l^{\mp}\stackrel{(-)}{\nu}
+ X$ processes. We find that the ability of probing the LED effects
can be improved by taking the cutoffs for the invariant mass of the
$W$-pair and the transverse momentum of the final lepton. Our
results demonstrate that the NLO QCD corrections to observables are
significant, and do not show any improvement for the
renormalization/factorization scale uncertainty on the QCD NLO
corrected cross section, because the leading-order result underestimates the
scale dependence.
\end{abstract}


\vfill \eject

\baselineskip=0.32in

\renewcommand{\theequation}{\arabic{section}.\arabic{equation}}
\renewcommand{\thesection}{\Roman{section}.}
\newcommand{\nb}{\nonumber}

\newcommand{\Dir}{\kern -6.4pt\Big{/}}
\newcommand{\Dirin}{\kern -10.4pt\Big{/}\kern 4.4pt}
\newcommand{\DDir}{\kern -7.6pt\Big{/}}
\newcommand{\DGir}{\kern -6.0pt\Big{/}}

\makeatletter      
\@addtoreset{equation}{section}
\makeatother       

\vskip 5mm
\section{Introduction}
\par
Motivated by the theoretical problems in the standard model (SM),
many extended models beyond the SM have been established. Among them the
large extra dimensions (LED) model proposed by Arkani-Hamed,
Dimopoulos, and Dvali in Ref.\cite{1} may be one of the promising
models which can solve the long-standing mass hierarchy problem.
This model used the idea of extra dimensions to bring gravity
effects from the Plank scale down to the electroweak scale. In the
LED model, the spacetime dimension is $D=4 + \delta$ with $\delta$
being the dimension of extra space, where the gravity and gauge
interactions are unified at one fundamental scale $M_S \sim TeV$
(the order of the electroweak scale). The graviton propagates in the
$D$-dimensional spacetime, while the SM particles exist only in the
usual ($3+1$)-dimensions.

\par
Taking into account of the bad behavior of quantum gravity in the
ultraviolet (UV) region, it is expedient to construct a low-energy
effective theory to describe the gravity-gauge-matter system in the
current (3+1)-dimensional spacetime. In the phenomenological sense,
this can be achieved through the Kaluza-Klein (KK) reduction in the
brane-world scenario \cite{2}. After applying this treatment to the
LED model, a $D$-dimensional massless graviton can be perceived as a
tower of massive KK modes propagating in the (3+1)-dimensional
spacetime. It turns out that the weakness of gravitational coupling
to the SM particles, suppressed by $\overline{M}_P$ (the reduced
Planck scale $\overline{M}_P=\frac{M_P}{\sqrt{8\pi}}$), can be
compensated by summing over numerous KK states. This scenario can
result in distinct effects at the high-energy colliders \cite{3}. Up
to now, many studies on the virtual KK graviton effects up to
the QCD next-to-leading order (NLO) in the LED model have
emerged. These include the processes of fermion-pair, multijet, and
vector-boson-pair production \cite{4,5,6}. In Ref.\cite{CMS-1} the CMS
Collaboration has performed a search for LED in the diphoton final
state events at the $\sqrt{s}= 7~TeV$ LHC with an integrated
luminosity of $36~pb^{-1}$. They set lower limits on the cutoff
scale $M_S$ in the range $1.6-2.3~TeV$ at the $95\%$ confidence
level. The dijet angular distribution results from the CMS and ATLAS
experiments appeared in Ref.\cite{CMS-2} and provide even stronger
limits on $M_S$, i.e., $M_S > 3.4~TeV$ (CMS) and $M_S > 3.2~TeV$
(ATLAS). Recently, the production of a $W$-pair at hadronic colliders
in the LED model has been studied up to the QCD NLO by Neelima
Agarwal, {\it et al} \cite{7}.

\par
In this paper, we revisit the NLO QCD corrections to the $W$-pair
production process at the LHC in the framework of the LED model, and
improve upon the results of Ref.\cite{7} by including the effects of
top-quark mass and the contribution from the $b\bar{b}$-fusion channel.
We provide the LED effect discovery and exclusion regions, the
kinematical distributions up to NLO in QCD by taking into account
the subsequential $W$-boson leptonic decay. The rest of the paper is
organized as follows. In Sec. II, we briefly go into the related
Feynman rules in the LED model. In Sec. III, the leading-order(LO) cross section for the \ppww process is described. In Sec. IV, we calculate the
NLO QCD corrections. In Sec. V, we present the numerical results
for the LO and NLO QCD corrected integrated cross section for the
$W$-pair production process and the distributions of final $W$-boson
decay products. Finally, a short summary is given.

\vskip 5mm
\section{ Related theories}
\label{related theories}
\par
The LED model consists of the pure gravity sector and the SM sector.
In this model the manifold, in which gravity propagates, is not the
ordinary four-dimensional spacetime manifold $\mathbb{R}^4$, but
$\mathbb{R}^4 \times {\cal M}$, where ${\cal M}$ is a compact
manifold of dimension $\delta$. For simplicity, one can tentatively
assume that ${\cal M}$ is a $\delta$-torus with radius $R$ and
volume $V_{\delta} = (2 \pi R)^{\delta}$ without loss of physical
significance.

\par
In our work we use the de Donder gauge. The Feynman rules for the
propagator of the spin-2 KK graviton and the relevant vertices which we
use are listed below. There $G_{\rm KK}^{\mu \nu}$, $\psi$, $W^{\pm
\mu}$, $A^{a \mu}$, and $\eta^a$ represent the fields of the
graviton, quark, $W$-boson, gluon, and $SU(3)$ ghost, respectively.
\begin{itemize}
\item
$\textrm{spin-2}~{\rm KK~graviton}~\textrm{propagator}~\textrm{
after summation over KK states}: $
\begin{eqnarray}
\tilde{G}_{\rm KK}^{\mu \nu \alpha \beta}=\frac{1}{2} D(s)
\left[\eta^{\mu \alpha} \eta^{\nu \beta} +
 \eta^{\mu \beta} \eta^{\nu \alpha} - \frac{2}{D-2}\eta^{\mu \nu} \eta^{\alpha \beta} \right]
\end{eqnarray}
\item
$G_{\rm KK}^{\mu
\nu}(k_3)-\bar{\psi}(k_1)-\psi(k_2)~\textrm{vertex}: $
\begin{eqnarray}
-i \frac{1}{4\overline{M}_P} \left[\gamma^{\mu} (k_1 + k_2)^{\nu} +
\gamma^{\nu} (k_1 + k_2)^{\mu} - 2 \eta^{\mu \nu} (\rlap/{k}_1 +
\rlap/{k}_2 - 2 m_{\psi}) \right]
\end{eqnarray}
\item
$G_{\rm KK}^{\mu
\nu}(k_4)-\bar{\psi}(k_1)-\psi(k_2)-A^{a\rho}(k_3)~\textrm{vertex}:
$
\begin{eqnarray}
i g_{s} \frac{1}{2\overline{M}_P} \left( \gamma^{\mu} \eta^{\nu
\rho} + \gamma^{\nu} \eta^{\mu \rho} - 2 \gamma^{\rho}\eta^{\mu \nu}
\right)T^{a}
\end{eqnarray}
\item
$G_{\rm KK}^{\mu
\nu}(k_3)-A^{a\rho}(k_1)-A^{b\sigma}(k_2)~\textrm{vertex}: $
\begin{eqnarray}
i \frac{2}{\overline{M}_P} \delta^{a b} \left[(C^{\mu \nu \rho
\sigma \tau \beta} - C^{\mu \nu \rho \beta \sigma \tau}) k_{1\tau}
k_{2\beta} + \frac{1}{\alpha_3}E^{\mu \nu \rho
\sigma}(k_1,k_2)\right]
\end{eqnarray}
\item
$G_{\rm KK}^{\mu \nu}(k_3)-W^{+ \rho}(k_1)-W^{-
\sigma}(k_2)~\textrm{vertex}: $
\begin{eqnarray}
i \frac{2}{\overline{M}_P} \left[B^{\mu \nu \rho \sigma} m_W^2 +
(C^{\mu \nu \rho \sigma \tau \beta} - C^{\mu \nu \rho \beta \sigma
\tau}) k_{1\tau} k_{2\beta} + \frac{1}{\xi}E^{\mu \nu \rho
\sigma}(k_1,k_2)\right]
\end{eqnarray}
\item
$G_{\rm KK}^{\mu
\nu}(k_4)-A^{a\rho}(k_1)-A^{b\sigma}(k_2)-A^{c\lambda}(k_3)~\textrm{vertex}:
$
\begin{eqnarray}
\frac{2}{\overline{M}_P}g_{s}f^{a b c} \left[(k_1-k_3)_{\tau}C^{\mu
\nu \tau \sigma \rho \lambda}+ (k_2-k_1)_{\tau}C^{\mu \nu \sigma
\rho \tau \lambda} + (k_3-k_2)_{\tau}C^{\mu \nu \lambda \sigma \tau
\rho}\right]
\end{eqnarray}
\item
$G_{\rm KK}^{\mu
\nu}(k_5)-A^{a\rho}(k_1)-A^{b\sigma}(k_2)-A^{c\lambda}(k_3)-A^{d\delta}(k_4)~\textrm{vetex}:
$
\begin{eqnarray}
-i \frac{1}{\overline{M}_P}g_{s}^2 [f^{e a c}f^{e b d}D^{\mu \nu
\rho \sigma \lambda \delta}+ f^{e a b}f^{e c d}D^{\mu \nu \rho
\lambda \sigma \delta}+ f^{e a d}f^{e b c}D^{\mu \nu \rho \sigma
\delta \lambda} ]
\end{eqnarray}
\item
$G_{\rm KK}^{\mu
\nu}(k_3)-\bar{\eta}^a(k_1)-{\eta}^b(k_2)~\textrm{vertex}: $
\begin{eqnarray}
- i \frac{2}{\overline{M}_P}\delta^{a b}B^{\alpha \beta \mu \nu}
k_{1\alpha} k_{2\beta}
\end{eqnarray}
\item
$G_{\rm KK}^{\mu
\nu}(k_3)-\bar{\eta}^a(k_1)-{\eta}^b(k_2)-A^{c\rho}(k_3)~\textrm{vertex}:
$
\begin{eqnarray}
\frac{2}{\overline{M}_P}g_{s}f^{a b c} B^{\alpha \rho \mu \nu}
k_{1\alpha}
\end{eqnarray}
\end{itemize}
where $g_s$ is the strong coupling constant, $T^a$ and $f^{abc}$ are
SU(3) generators and structure constants, $D = n + \delta$, $n = 4 -
2 \epsilon$, $\overline{M}_p$ is the reduced Planck mass, $\alpha_3$ and
$\xi$ are SU(3) and charged SU(2) gauge fixing parameters, and
$D(s)$ can be expressed as \cite{2}
\begin{equation}
D(s)\ =\ {s^{\delta/2-1}\over\Gamma(\delta/2)}
{R^{\delta}\over(4\pi)^{\delta/2}} \biggl[\pi + 2i
I(\Lambda/\sqrt{s})\biggr] \label{DS}
\end{equation}
and
\begin{equation}
I(\Lambda/\sqrt{s})\ =\ P \int_0^{\Lambda/\sqrt{s}}dy\
{y^{\delta-1}\over 1-y^2}\ . \label{B6}
\end{equation}
The integral $I(\Lambda/\sqrt{s})$ contains an ultraviolet cutoff
$\Lambda$ on the KK modes \cite{2,3}. In this work we set it to
be the fundamental scale $M_S$. It should be understood that a
point $y=1$ has been removed from the integration path. Besides,
all the momenta are assumed to be incoming to the vertices,
except that the fermionic momenta are set to be along the fermion
flow directions. The coefficients $A^{\mu \nu}$, $B^{\mu \nu
\alpha \beta}$, $C^{\rho \sigma \mu \mu \alpha \beta}$,
$D^{\mu \nu \rho \sigma \lambda \delta}$, and
$E^{\mu \nu \rho \sigma}(k_{1},k_{2})$ are expressed as
\begin{eqnarray}
A^{\mu \nu} & = & \frac{1}{2}\eta^{\mu \nu},~~~~~~~ B^{\mu \nu
\alpha \beta} = \frac{1}{2}
      (\eta^{\mu \nu}\eta^{\alpha \beta}
      -\eta^{\mu \alpha}\eta^{\nu \beta}
      -\eta^{\mu \beta}\eta^{\nu \alpha}),
       \nb \\
C^{\rho \sigma \mu \nu \alpha \beta} & = & \frac{1}{2}
      [\eta^{\rho \sigma}\eta^{\mu \nu}\eta^{\alpha \beta}
     -(\eta^{\rho \mu}\eta^{\sigma \nu}\eta^{\alpha \beta}
      +\eta^{\rho \nu}\eta^{\sigma \mu}\eta^{\alpha \beta}
      +\eta^{\rho \alpha}\eta^{\sigma \beta}\eta^{\mu \nu}
      +\eta^{\rho \beta}\eta^{\sigma \alpha}\eta^{\mu \nu})],
      \nb \\
D^{\mu \nu \rho \sigma \lambda \delta} & = &
      \eta^{\mu \nu}(\eta^{\rho \sigma}\eta^{ \lambda \delta}
      -\eta^{\rho \delta}\eta^{ \sigma \lambda})
     +(\eta^{\mu \rho}\eta^{\nu \delta}\eta^{\lambda \sigma}
      +\eta^{\mu \lambda}\eta^{\nu \sigma}\eta^{\rho \delta}
        -\eta^{\mu \rho}\eta^{\nu \sigma}\eta^{\lambda \delta}
      -\eta^{\mu \lambda}\eta^{\nu \delta}\eta^{\rho \sigma}
      + (\mu  \leftrightarrow \nu)), \nb \\
E^{\mu \nu \rho \sigma}(k_{1},k_{2}) & = &
      \eta^{\mu \nu}(k_1^{\rho} k_1^{\sigma} + k_2^{\rho} k_2^{\sigma}
      + k_1^{\rho} k_2^{\sigma}) - \left [\eta^{\nu \sigma} k_1^{\mu} k_1^{\rho}
      + \eta^{\nu \rho} k_2^{\mu} k_2^{\sigma} + (\mu \leftrightarrow \nu)\right ]. \nb
\end{eqnarray}

\par
We code programmatically the related Feynman rules in the
FeynArts 3.5 package \cite{10} to generate the Feynman diagrams and
the relevant amplitudes. The FormCalc 5.4 \cite{11} package is
implemented subsequently to simplify the amplitudes.

\vskip 5mm
\section{ LO cross section for \ppww }
\par
\begin{figure*}
\begin{center}
\includegraphics[scale=0.8]{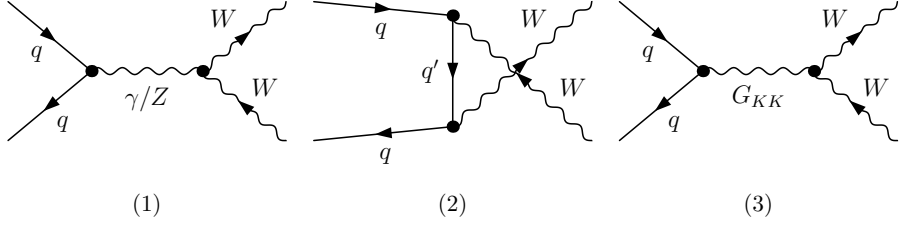}
\caption{\label{fig1} The tree-level Feynman diagrams for the
partonic processes \qqww in the LED model. (1) and (2) are the
SM-like diagrams, where $q$ represents the $u$-, $d$-, $c$-, $s$- and
$b$-quark. (3) is the extra diagram with KK graviton exchange. }
\end{center}
\end{figure*}
\begin{figure*}
\begin{center}
\includegraphics[scale=0.8]{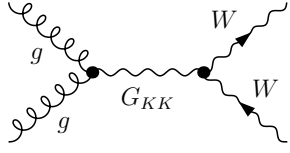}
\caption{\label{fig2} The tree-level Feynman diagram for the
partonic process \ggww in the LED model. }
\end{center}
\end{figure*}

We treat the up-, down-, charm-, strange-, and bottom-quark as massless
particles, and adopt the five-flavor scheme in the leading
order and QCD next-to-leading order calculations. The LO
contribution to the parent process \ppww includes the
quark-antiquark $(q=u,d,s,c,b)$ annihilations and the gluon-gluon
fusion partonic processes: $q(p_1)+\bar q(p_2) \to
W^+(p_3)+W^-(p_4)$ and $g(p_1)+g(p_2) \to W^+(p_3)+W^-(p_4)$. There
$p_{i}$ $(i=1,2,3,4)$ represent the four-momenta of the incoming and
outgoing particles, respectively. The corresponding Feynman diagrams
are shown in Figs.\ref{fig1} and \ref{fig2}. Figures.\ref{fig1}(1)
and .\ref{fig1}(2) are the LO SM-like diagrams for partonic process $q\bar q
\to W^+W^-$. In Fig.\ref{fig1}(1) the internal wavy line means
exchanging $\gamma$ or a $Z^0$-boson. There we ignore the diagrams
with exchanging Higgs boson, since the initial quarks are all
massless.

\par
We express the tree-level amplitudes for the partonic processes
$q\bar{q} \to W^+W^-$ and $gg \to W^+W^-$ as
\begin{equation}
\label{TreeAmplitude1} {\cal M}_{q\bar{q}}^{0} = {\cal
M}_{q\bar{q}}^{0,SM} + {\cal M}_{q\bar{q}}^{0,LED},~~~~{\cal
M}_{gg}^{0} = {\cal M}_{gg}^{0,LED},
\end{equation}
where ${\cal M}_{q\bar{q}}^{0,SM}$ $(q=u,d,c,s,b)$ is the amplitude
contributed by the tree-level SM-like diagrams, while ${\cal
M}_{q\bar{q}}^{0,LED}$ and ${\cal M}_{gg}^{0,LED}$ are the
tree-level amplitudes with KK graviton exchange. Our calculations
show that the analytical expression of the SM matrix element squares
summed (averaged) over the final (initial) state spins and colors at
the LO, $\overline{|{\cal M}_{q\bar{q}}^{0,SM}|^2}$, for the
partonic process without massive internal or external quark (i.e., $q
= u, d, c, s$), is the same as that presented in Ref.\cite{7}. But
for the partonic process $b \bar{b} \to W^+ W^-$, there is a
t-channel diagram with a massive top-quark exchange. Its explicit
expression of $\overline{|{\cal M}_{b\bar{b}}^{0,SM}|^2}$ is
presented below by adopting the notations in Ref.\cite{7}.
\begin{equation}
\label{Amplitude-t} \overline{|{\cal M}_{b\bar{b}}^{0,SM}|^2}=
\frac{e^4}{N} \left(A_1^b B_{1m}^b +A_2^b B_{2m}^b +A_3^b
B_{3m}^b\right),
\end{equation}
where N is the number of colors. The explicit expressions for
$A_1^b$, $A_2^b$, and $A_3^b$ can be obtained from the Eqs.(8) in
Ref.\cite{7} by taking the replacement of $u \to b$. The kinematic
invariants $B_{1m}^b$, $B_{2m}^b$, and $B_{3m}^b$ can be expressed as
\begin{equation}
B_{1m}^b = \frac{u^2}{(u-m_t^2)^2}B_1^d(t,u,s),~~~ B_{2m}^b =
B_2^d(t,u,s),~~~ B_{3m}^b = \frac{u}{(u-m_t^2)}B_3^d(t,u,s),
\end{equation}
where $B_1^d(t,u,s)$, $B_2^d(t,u,s)$ and $B_3^d(t,u,s)$ are
presented in Eqs.(9)-(12) of Ref.\cite{7}. Then the LO cross
sections for the unpolarized $W$-pair production processes at the
partonic level can be expressed as
\begin{eqnarray}\label{int}
\hat{\sigma}_{ij}^{LO} &=& \frac{1}{4|\vec{p}|\sqrt{\hat{s}}}\int
{\rm d}\Gamma_2 \overline{|{\cal M}_{ij}^{0}|^2},
~~(ij=u\bar{u},d\bar{d},c\bar{c},s\bar{s},b\bar{b},gg),
\end{eqnarray}
where $\vec{p}$ is the momentum of one initial parton in
center-of-mass system (c.m.s.) and ${\rm d} \Gamma_2$ is the two-body
phase space element expressed as
\begin{eqnarray}
{\rm d} \Gamma_2 = (2\pi)^4  \delta^{(4)} (p_1+p_2-p_3 - p_4)
\prod_{i=3,4} \frac{d^3 \vec{p}_i}{(2\pi)^3 2E_i}.
\end{eqnarray}

\par
By convoluting $\hat{\sigma}^{LO}_{i j}$  with the parton
distribution functions (PDFs) of the colliding protons, the LO cross
section for the parent process, \ppww, can be written as
\begin{eqnarray}
\sigma_{LO} &=& \sum^{c\bar c,b\bar b,gg}_{ij=u\bar u,d\bar d,s\bar
s}\frac{1}{1+\delta_{ij}} \int dx_A dx_B \left[ G_{i/A}(x_A,\mu_f)
G_{j/B}(x_B,\mu_f)\hat{\sigma}^{LO}_{ij}(\sqrt{\hat{s}})
+(i \leftrightarrow j) \right], \nb \\
\end{eqnarray}
where $G_{i/P}$ $(i = g, q, \bar{q})$ represent the PDFs of parton
$i$ in proton $P$, $\mu_{f}$ is the factorization scale,
$\sqrt{\hat{s}}=x_A x_B \sqrt{s}$, $x_A$ and $x_B$ describe the
momentum fractions of parton (gluon or quark) in protons $A$ and
$B$, respectively.

\vskip 5mm
\section{NLO QCD corrections }
\par
The complete NLO QCD correction to the parent process \ppww consists
of following components. (1) The virtual contribution from the QCD
one-loop and the corresponding counterterm diagrams to the partonic
channels $q\bar{q} \to W^+W^-$ and $gg \to W^+W^-$. (2) The
contribution of the real gluon emission partonic processes. (3) The
contribution of the real light-(anti)quark emission partonic
processes. And (4) the corresponding contribution of the PDF
counterterms. There inevitably exist the ultraviolet (UV) and
infrared (IR) divergences in the NLO calculations, and we adopt the
dimensional regularization scheme in $n=4-2 \epsilon$
dimensions to isolate and manipulate these divergences.

\par
{A. Virtual corrections }
\par
\begin{figure*}
\includegraphics[scale=0.8]{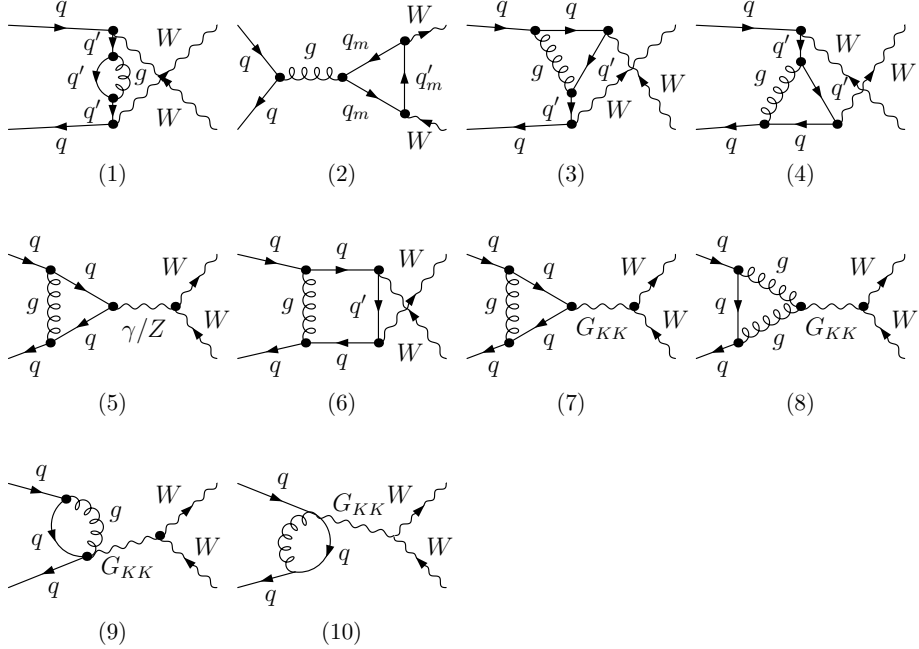}
\vspace*{-0.3cm} \centering \caption{\label{fig3} The QCD one-loop
Feynman diagrams for the partonic process \qqww. (1)-(6) are the
SM-like diagrams. (7)-(10) are the diagrams with KK graviton
exchange.}
\end{figure*}
\begin{figure*}
\begin{center}
\includegraphics[scale=0.8]{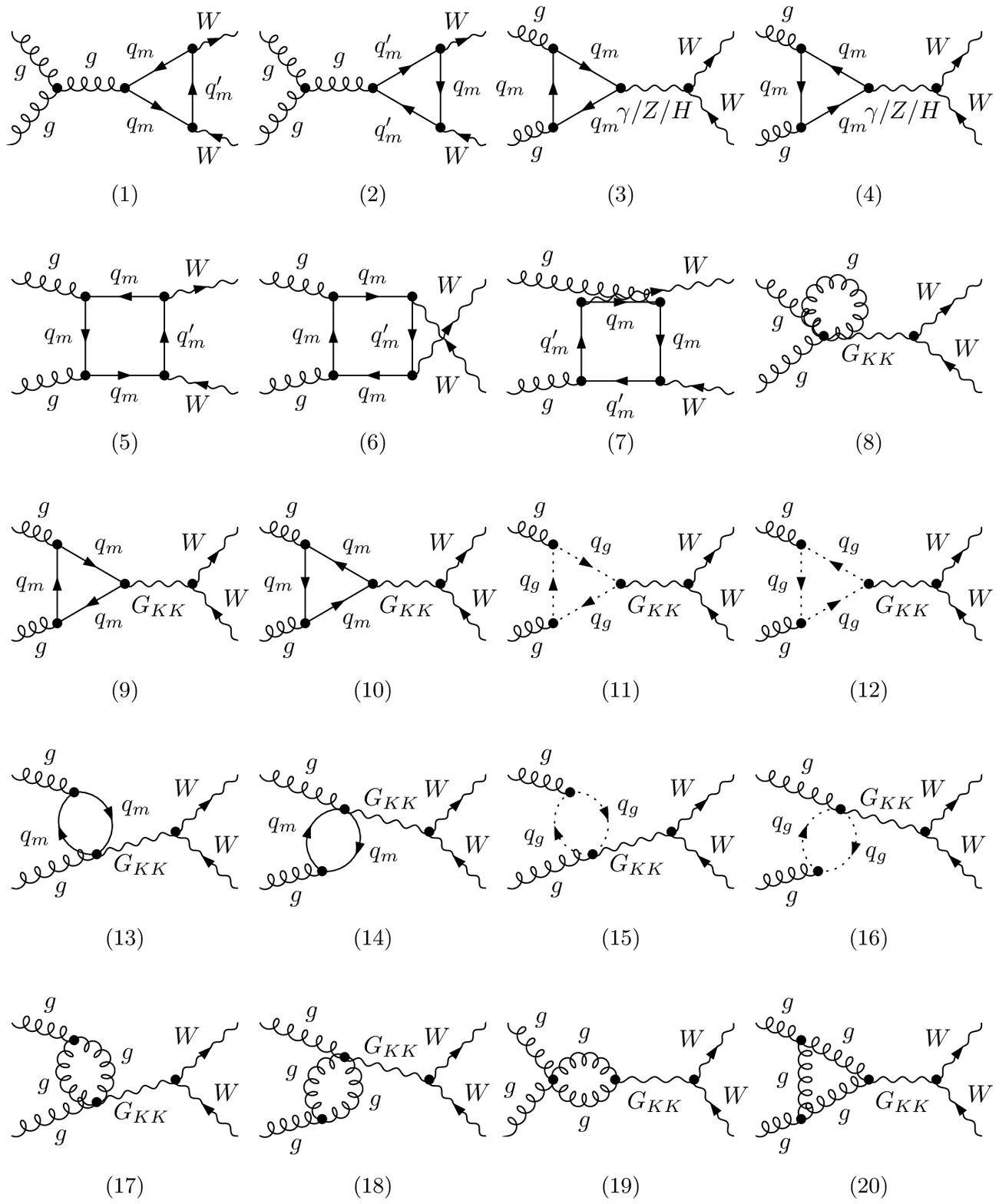}
\caption{\label{fig4} The QCD  one-loop Feynman diagrams for
partonic process \ggww. (1)-(7) are the SM-like diagrams. (8)-(20) are
the diagrams with KK graviton exchange. In all diagrams $q_m$
represents $u$-, $d$-, $c$-, $s$-, $b$- and $t$-quark except the
diagrams in Figs.\ref{fig4}(3) and Fig.\ref{fig4}(4) involving the
coupling between Higgs boson and top quarks, where $q_m$ denotes
only top quark. }
\end{center}
\end{figure*}

The Feynman diagrams for the virtual corrections to the \qqww and
\ggww partonic processes are shown in Fig.\ref{fig3} and
Fig.\ref{fig4}, respectively. In Figs.\ref{fig4}(3) and .\ref{fig4}(4) the
diagrams involving Yukawa coupling between Higgs boson and top
quarks are included, but the diagrams involving Yukawa coupling
between Higgs boson and massless quarks are excluded due to their
vanishing contribution. There exist UV and soft/collinear IR
singularities in the calculations of these one-loop diagrams. To
remove the UV divergences, we need only the wave function
renormalization constants for the quark and gluon fields. We
introduce the renormalization constants $\delta Z_{\psi_{q,L,R}}$
for massless quark (q=u,d,c,s,b) fields and $\delta Z_{A}$ for the gluon
field defined as
\begin{eqnarray}
\psi^{0}_{q,L,R} = (1+\delta Z_{\psi_{q,L,R}})^{1/2}\psi_{q,L,R},~~~
A^{a0}_{\mu} = (1+\delta Z_{A})^{1/2}A^{a}_{\mu}.
\end{eqnarray}
In the modified minimal subtraction ($\overline{MS}$) renormalization scheme the renormalization
constants for the massless quarks are expressed as
\begin{eqnarray}
\delta Z_{\psi_{q,L}} &=&
-\frac{\alpha_{s}}{4\pi}C_{F}(\Delta_{UV}-\Delta_{IR}),
~~~\delta Z_{\psi_{q,R}} = -\frac{\alpha_{s}}{4\pi}C_{F}\left(\Delta_{UV}-\Delta_{IR}\right),  \\
\delta Z_{A} &=&
\frac{\alpha_{s}}{4\pi}\left(\frac{5}{3}C_{A}-\frac{4}{3}n^{UV}_{f}T_{F}\right)\Delta_{UV}
+\frac{\alpha_{s}}{4\pi}\left(\frac{5}{3}C_{A}-\frac{4}{3}n^{IR}_{f}T_{F}\right)\Delta_{IR},
\end{eqnarray}
To remove the UV and IR divergences in the $b\bar b$-fusion subprocess, we need
introduce the counterterms for the top-quark field and its mass, i.e.,
\begin{eqnarray}
\psi^{0}_{t,L,R} = (1+\delta Z_{\psi_{t,L,R}})^{1/2}\psi_{t,L,R},~~~
m^{0}_{t} = m_{t} + \delta m_{t}.
\end{eqnarray}
We use the on-mass-shell scheme to renormalize the top-quark field and
mass. They are expressed as
\begin{eqnarray}
\delta Z_{\psi_{t,L}} &=&
-\frac{\alpha_{s}}{4\pi}C_{F}(\Delta_{UV}+2\Delta_{IR}+3\ln{\frac{\mu_r^{2}}{m^{2}_{t}}}+4), \\
\delta Z_{\psi_{t,R}} &=&
-\frac{\alpha_{s}}{4\pi}C_{F}(\Delta_{UV}+2\Delta_{IR}+3\ln{\frac{\mu_r^{2}}{m^{2}_{t}}}+4),  \\
\frac{\delta m_{t}}{m_t} &=&
-\frac{3\alpha_{s}}{4\pi}C_{F}(\Delta_{UV}+\ln{\frac{\mu_r^{2}}{m^{2}_{t}}}+\frac{4}{3}),
\end{eqnarray}
In the above equations $\mu_r$ is the renormalization scale,
$C_{F}=\frac{4}{3}$, $C_{A}=3$, $T_{F}=\frac{1}{2}$, $n^{UV}_{f}=6$
corresponds to the six flavor quarks ($u$, $d$, $c$, $s$, $t$, $b$),
whereas $n^{IR}_{f}=5$ is the number of the massless quarks ($u$,
$d$, $s$, $c$, $b$). Moreover, $\Delta_{UV}=
\frac{1}{\epsilon_{UV}}\Gamma
(1+\epsilon_{UV})(4\pi)^{\epsilon_{UV}}$ and
$\Delta_{IR}=\frac{1}{\epsilon_{IR}}\Gamma
(1+\epsilon_{IR})(4\pi)^{\epsilon_{IR}}$ refer to the UV and IR
divergences, respectively.

\par
Then the results for the differential cross sections for the $q\bar
q$ annihilation and $gg$ fusion partonic channels are UV finite but
soft/collinear IR divergent. The soft/collinear IR singularities can
be canceled by adding the contributions of the real emission
partonic processes and the corresponding PDF counterterms.

\par
{B. Real gluon emission}
\par
\begin{figure*}
\begin{center}
\includegraphics [scale=0.8]{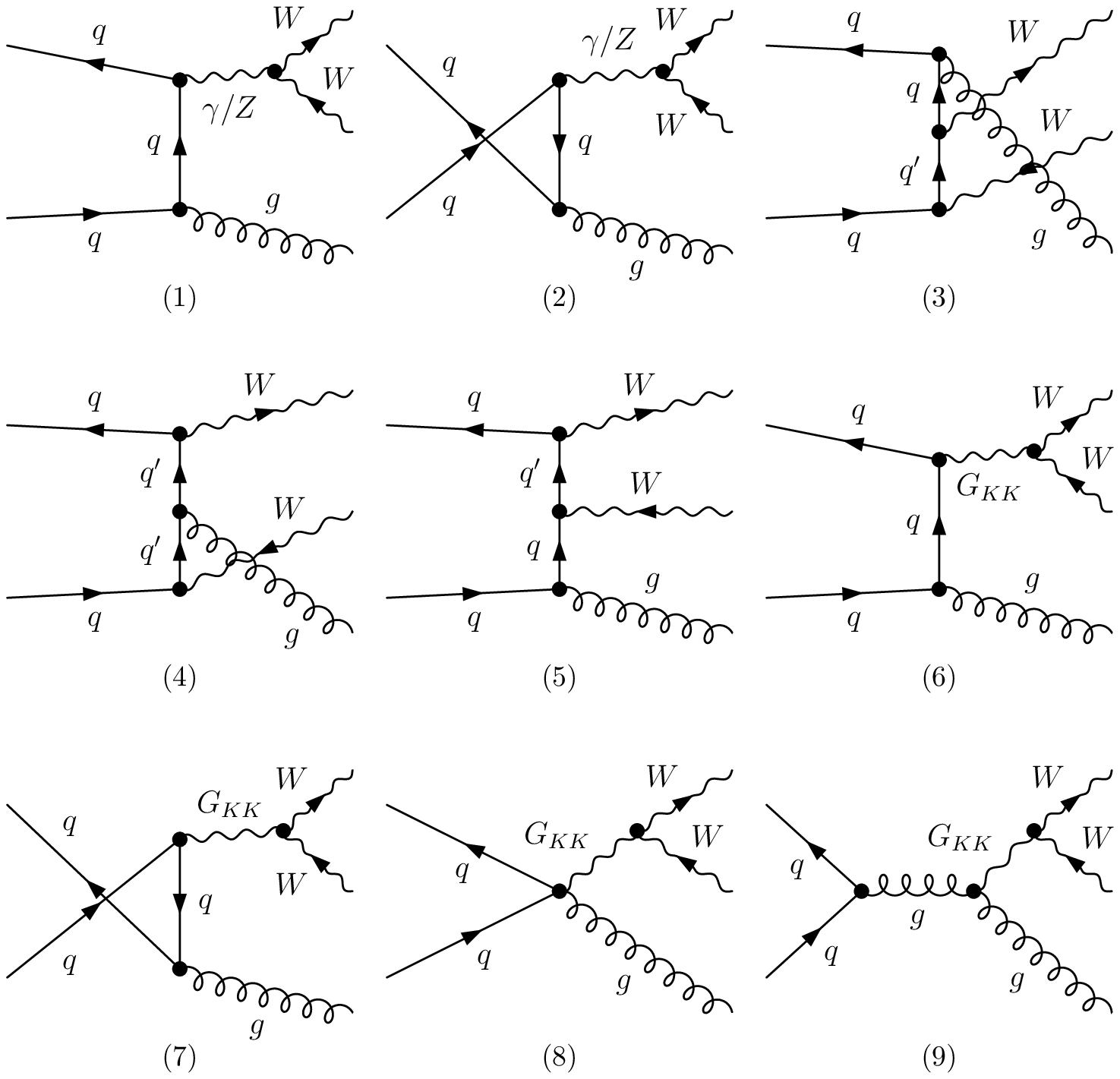}
\caption{\label{fig5} The tree-level Feynman diagrams for the real
gluon emission subprocess \qqwwg. (1)-(5) are the SM-like diagrams.
(6)-(9) are the extra diagrams with KK graviton exchange.}
\end{center}
\end{figure*}
\begin{figure*}
\begin{center}
\includegraphics [scale=0.8]{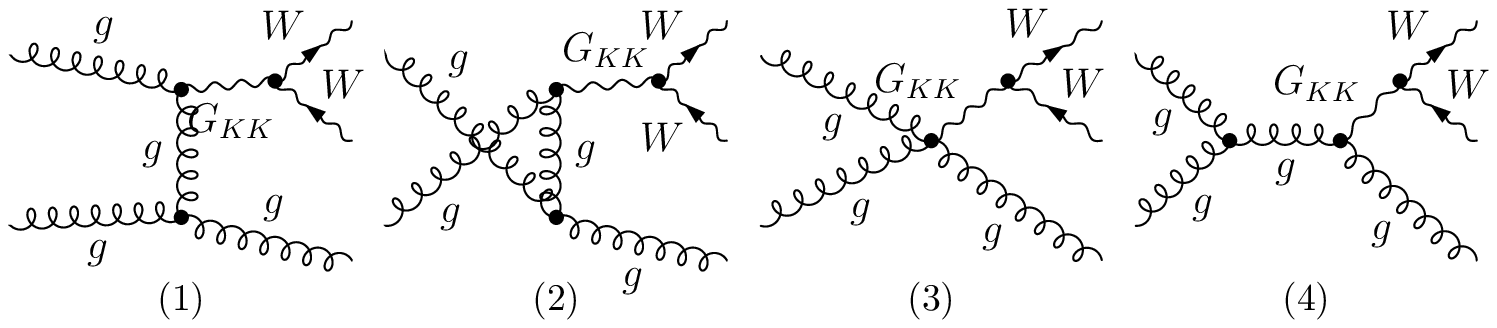}
\caption{\label{fig6} The tree-level Feynman diagrams for the real
gluon emission subprocess \ggwwg. There is no SM-like diagram.}
\end{center}
\end{figure*}

The real gluon emission contributions are from $g(p_1)+g(p_2) \to
W^+(p_3) + W^-(p_4) + g(p_5)$ and $q(p_1)+\bar q (p_2) \to W^+(p_3)
+ W^-(p_4) + g(p_5)$ partonic processes. The corresponding Feynman
diagrams are shown in Fig.\ref{fig5} and Fig.\ref{fig6},
respectively. We employ the two cutoff phase space slicing (TCPSS)
method \cite{13} to calculate the contributions from the real gluon
emission partonic processes. An arbitrary soft cutoff $\delta_{s}$
is introduced to separate the gluon emission subprocess phase space
into two regions, soft gluon and hard gluon regions. Furthermore,
another cutoff $\delta_{c}$ is introduced to decompose the real hard
gluon emission phase space region into hard collinear ($HC$)
and hard noncollinear ($\overline{HC}$) regions. The partonic
differential cross section for the real gluon emission subprocess
can be expressed as
\begin{eqnarray}
d\hat{\sigma}_g ~=~ d\hat{\sigma}^S_g+d\hat{\sigma}^H_g &=&
d\hat{\sigma}^S_g+d\hat{\sigma}^{HC}_g+d\hat{\sigma}^{\overline{HC}}_g.
\end{eqnarray}

\par
{C. Real light-(anti)quark emission}
\par
\begin{figure*}
\begin{center}
\includegraphics [scale=0.8]{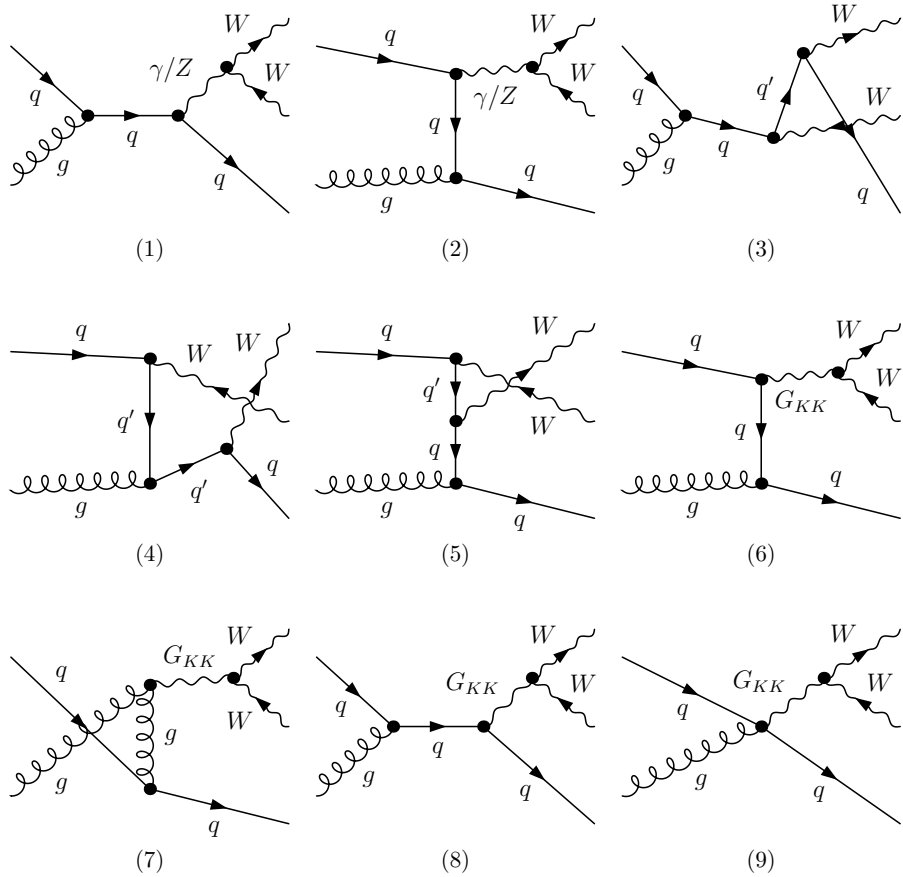}
\caption{\label{fig7} The tree-level Feynman diagrams for the real
light-(anti)quark emission subprocesses \qgwwq. (1)-(5) are the
SM-like diagrams. (6)-(9) are the diagrams with KK graviton exchange.}
\end{center}
\end{figure*}

In addition to the real gluon emission discussed above, there are
contributions from the massless light-(anti)quark
($u,d,c,s,b$,$\bar{u},\bar{d},\bar{c},\bar{s},\bar{b}$) emission
partonic processes. In the five-flavor scheme the massless
light-quark $q$ involves $u$-, $d$-, $c$-, $s$-, $b$-quarks.
Considering the fact that the final (anti)bottom-quark can be tagged
in experiments and the collinear IR singularities of the real
(anti)bottom-quark emission subprocesses are completely canceled by
those of the corresponding PDF counter terms, we do not include the
contributions of the bottom and antibottom emissions, and adopt the
five-flavor PDFs \cite{14}. We depict the Feynman diagrams for the
partonic processes $qg  \to W^+  W^- + q$ and $\bar qg \to W^+  W^-
+ \bar q$ in Fig.\ref{fig7}. These partonic processes contain only
the initial state collinear singularities. By using the TCPSS method
described above, we can split the phase space into collinear and
noncollinear regions. The differential cross sections for the
partonic processes $qg \to W^+W^-q$ and $\bar{q}g \to W^+W^-\bar{q}$
can then be expressed as
\begin{eqnarray}
d\hat{\sigma}_{q(\bar{q})}(q(\bar{q}) g \to W^+W^-q(\bar{q})) ~~=~~
d\hat{\sigma}^{C}_{q(\bar{q})} +
d\hat{\sigma}^{\overline{C}}_{q(\bar{q})},
\end{eqnarray}
where $d\hat{\sigma}^{\overline{C}}_{q}$ and
$d\hat{\sigma}^{\overline{C}}_{\bar{q}}$ are finite and can be
evaluated by using the general Monte Carlo method.

\par
{D. NLO QCD corrections to the \ppww process }
\par
Combining the renormalized virtual corrections and the real
gluon/light-(anti)quark emission contributions, the partonic cross
sections still contain the collinear divergence, which can be
absorbed into the redefinition of the PDFs at the NLO according to
the mass factorization \cite{15}. We find that after the summation
of all the NLO QCD corrections, the soft and collinear IR
divergences vanish. We can see from above discussion that the
final total ${\cal O}(\alpha_{s})$ corrections consist of the
two-body term $\sigma^{(2)}$ and the three-body term $\sigma^{(3)}$.
The total cross section up to the QCD NLO is expressed as
\begin{eqnarray}
\sigma_{NLO} = \sigma_{LO} + \Delta \sigma_{NLO}= \sigma_{LO} + \sigma^{(2)} + \sigma^{(3)}.
\end{eqnarray}
It is UV finite, IR safe, and cutoff $\delta_{c}/\delta_{s}$
independent \cite{13,16}, which will further be checked in the
numerical evaluations.

\vskip 10mm
\section{Numerical results and discussions }
\par
In this section, we present the numerical results of the integrated
cross sections and the kinematic distributions of the final
particles for the \ppww process in both the SM and LED model up to
the QCD NLO. In order to verify the correctness of our numerical
calculations, we made the following checks:
\begin{itemize}
\item[(i)] In Table \ref{tab1}, we list our numerical results of the
LO and NLO QCD corrected integrated cross sections in the SM for the
\ppww process by taking the input parameters, PDFs, and event
selection criterion from Table 4 of Ref.\cite{17}. It
shows that our LO and NLO QCD corrected cross sections in the SM are
in good agreement with those in Ref.\cite{17} within the integration
errors.
\begin{table}[htbp]
\begin{center}
\begin{tabular}{|c|c|c|c|}
\hline
{} LHC          &Ref.\cite{17} &FeynArts           &CompHEP          \\
\hline  Born[pb]        &86.7                &86.711(6)      &86.7(1)         \\
\hline  NLO QCD[pb]     &127.8               &127.7(1)       &-----           \\
\hline
\end{tabular}
\caption{  \label{tab1} The LO and NLO QCD corrected cross sections
for the $pp \to W^{+}W^{-}+X$ process in the SM by taking the same
input parameters and event selection criterion as those in
Ref.\cite{17}. }
\end{center}
\end{table}

\item[(ii)] The UV and IR safeties are verified numerically after combining
all the contributions at the QCD NLO.

\item[(iii)] We calculate the NLO QCD corrections to integrated cross section for
the $pp \to u\bar{u} \to W^+W^- + X$ process as functions of the cutoff
$\delta_{s}$ at the $\sqrt{s} = 14~TeV$ LHC in the LED model, where we take
$\mu_f = \mu_r = \mu_0 = m_W$, $M_S = 3.5~TeV$, $\delta=3$ and
$\delta_{c} = \delta_s/50$. Some of the results are listed in Table \ref{tab1-1}.
It is shown clearly that the NLO QCD correction ($\Delta\sigma_{NLO}^{LED}$)
does not depend on the arbitrarily chosen values of $\delta_{s}$ and $\delta_{c}$
within the calculation errors. In the further numerical calculations,
we fix $\delta_s=10^{-3}$ and $\delta_c=\delta_s/50$.
\begin{table}[htbp]
\begin{center}
\begin{tabular}{|c|c|}
\hline
{} $\delta_{s}$        & $\Delta\sigma_{NLO}^{LED}$[pb]   \\
\hline  $2\times10^{-3}$ &   13.19(3)         \\
\hline  $1\times10^{-3}$ &   13.20(3)         \\
\hline  $7\times10^{-4}$ &   13.21(3)         \\
\hline  $4\times10^{-4}$ &   13.22(5)         \\
\hline  $2\times10^{-4}$ &   13.24(5)         \\
\hline  $1\times10^{-4}$ &   13.26(6)         \\
\hline  $7\times10^{-5}$ &   13.25(6)         \\
\hline  $4\times10^{-5}$ &   13.27(6)         \\
\hline  $2\times10^{-5}$ &   13.26(7)         \\
\hline
\end{tabular}
\caption{ \label{tab1-1} The dependence of the NLO QCD correction to
the integrated cross section for the process $pp \to u\bar{u} \to W^+W^- + X$
at the $\sqrt{s} = 14~TeV$ LHC in the LED model, where we set $\mu_f =
\mu_r = \mu_0 = m_W$, $M_S = 3.5~TeV$, $\delta=3$ and $\delta_{c} =
\delta_s/50$. }
\end{center}
\end{table}

\item[(iv)] We calculate the SM LO $W$-pair invariant mass
distribution ($d\sigma^{SM}_{LO}/dM_{WW}$) for the \ppww process
with the same input parameters, PDFs and event selection criterion
as those used in Ref.\cite{7}. The numerical results, which are
obtained by using FeynArts and CompHEP packages separately, are
coincident with each other.

\item[(v)] For further comparison with the previous work of N. Agarwal,
{\it et al}, we recalculate the LO and NLO QCD corrected
distributions of $W$-pair invariant mass in both the SM and LED
model ($d\sigma^{SM,LED}_{LO}/dM_{WW}$,
$d\sigma^{SM,LED}_{NLO}/dM_{WW}$) for the \ppww process at the
$\sqrt{s}=14~TeV$ LHC, where we set all the quarks being massless
except $m_t=172.0~GeV$, and take the PDFs and event selection
criterion from Ref.\cite{7}. We plot our LO and
NLO QCD corrected results in the SM in Fig.\ref{fig7-1}(a), and the
results in the LED model in Fig.\ref{fig7-1}(b). In these two
figures we depict also the corresponding curves from N. Agarwal's
paper \cite{7} for comparison. We can see that there exist obvious
discrepancies between ours and the corresponding N. Agarwal
results, especially in the large $M_{WW}$ region. One of the reasons
for the disagreement is because we have included the effects of
top-quark mass in our calculations. From Figs.\ref{fig7-1}(a,b) we
can see the K-factors of the QCD corrections increase to large
numbers of $K=5.29$ and $K=2.33$ separately, when $W$-pair invariant
mass approaches $M_{WW} = 1300~GeV$. This occurs because the
K-factors in Figs.\ref{fig7-1}(a,b) are the results with only
constraint on W-bosons ($|y_W| < 2.5$ ) \cite{7}. In this case the
differential cross section, $d\sigma^{SM}_{NLO}/dM_{WW}$, receives
a large contribution from the hard jet emission corrections
($W^+W^-+jet$). For example, in Fig.\ref{fig7-1}(a) at $M_{WW} =
1300~GeV$ the K-factor contributed by hard jet emission processes
can reach $3.59$.
\begin{figure}[htbp]
\begin{center}
\includegraphics[scale=0.7]{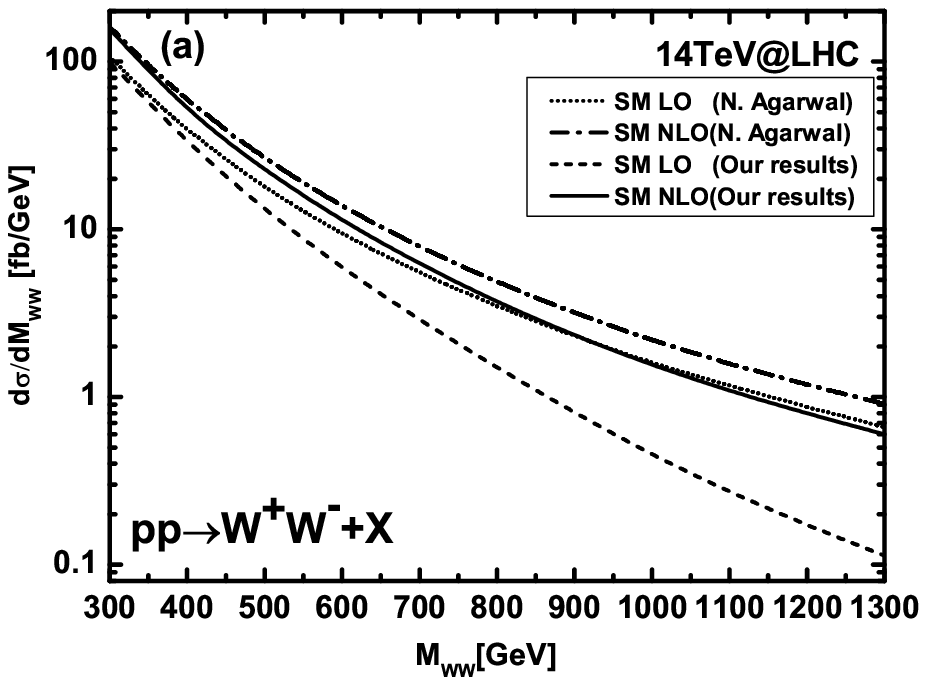}%
\hspace{0in}%
\includegraphics[scale=0.7]{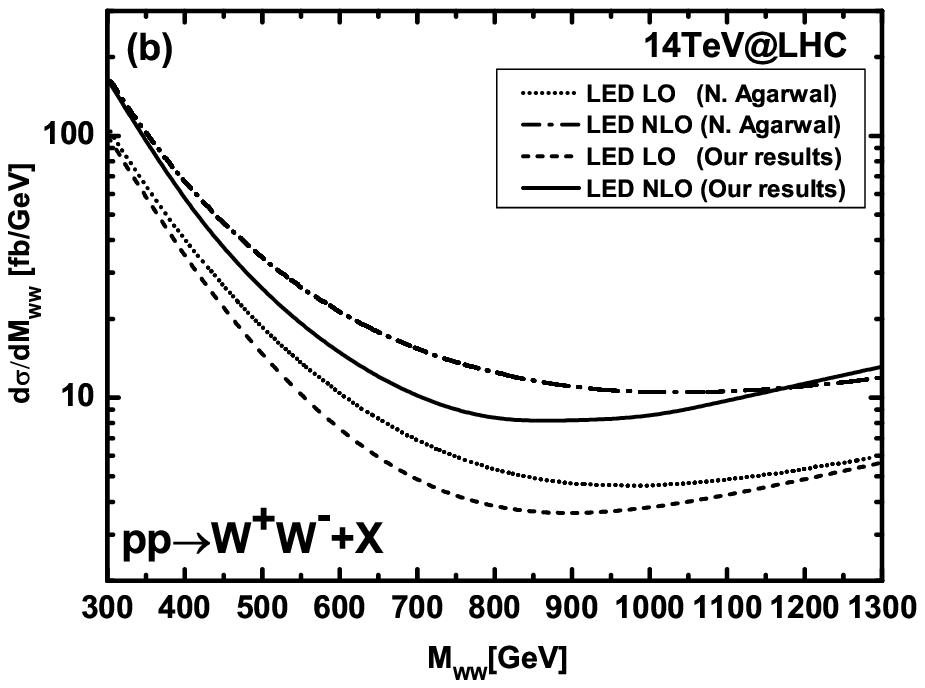}%
\hspace{0in}%
\caption{ \label{fig7-1} Invariant mass ($M_{WW}$) distributions at
the LO and NLO for the $pp \to W^{+}W^{-}+X$ process at the
$\sqrt{s}=14~TeV$ LHC. There the input parameters, PDFs, and event
selection criterion are taken from Ref.\cite{7} (where $M_S=2~TeV$,
$\delta=3$, and $\mu_r=\mu_f=M_{WW}$). For comparison with previous
work, we depict also the corresponding curves from N. Agarwal's
paper \cite{7} in both panels. (a) The distributions in the SM. (b)
The distributions in the LED model. }
\end{center}
\end{figure}
\end{itemize}

\par
In the following calculations we take one-loop and two-loop running
$\alpha_{s}$ in the LO and QCD NLO calculations, respectively
\cite{18}. The QCD parameters are taken as $\Lambda_5^{LO}=165~MeV$,
$\Lambda_5^{\overline{MS}}=226~MeV$, the number of the active
flavors is $n_{f}=5$, the Cabibbo-Kobayashi-Maskawa matrix is set as a unit matrix, and
the colliding c.m.s. energy is taken as $\sqrt{s}=7~TeV$ and
$\sqrt{s}=14~TeV$ for the early and future LHC. To satisfy the
unitary constraint, we adopt the cut $\sqrt{\hat{s}}<M_{S}$ for the
whole phase space. We assume $m_H=120~GeV$ and the renormalization
and factorization scales to be equal ($\mu_r=\mu_f \equiv \mu$), and
we define $\mu_0 \equiv m_W$. We use the CTEQ6L1 and CTEQ6M PDFs
\cite{19,20} in the LO and QCD NLO calculations, respectively. The
other related input parameters are taken from \cite{18}:
$\alpha^{-1}(m_{Z}) = 127.916$, $m_W = 80.399~GeV$,
$m_Z=91.1876~GeV$ and $m_t=172.0~GeV$. As we know that the constraints on the
final particles are necessary in realistic experimental
event collections, and the theoretical calculation should keep the perturbative
convergence. We adopt the following event selection constraints additionally.
(1) For the real emission contributions, we accept the events which satisfy
the condition that the jet pseudorapidity $|y_{jet}|>2.5$ or the jet transverse
momentum $p_{T}^{jet}<50~GeV$. (2) The $W$-pair invariant mass is restricted
in the range of $M_{WW}>400~GeV$.

\par
\begin{figure}[htbp]
\begin{center}
\includegraphics[scale=0.8]{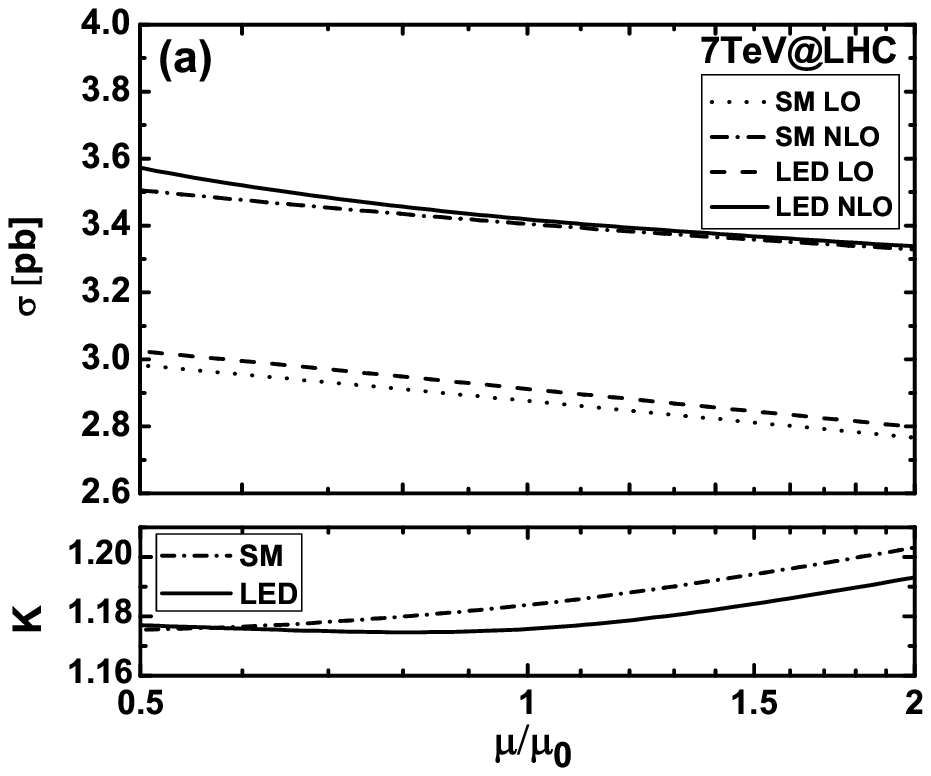}%
\includegraphics[scale=0.8]{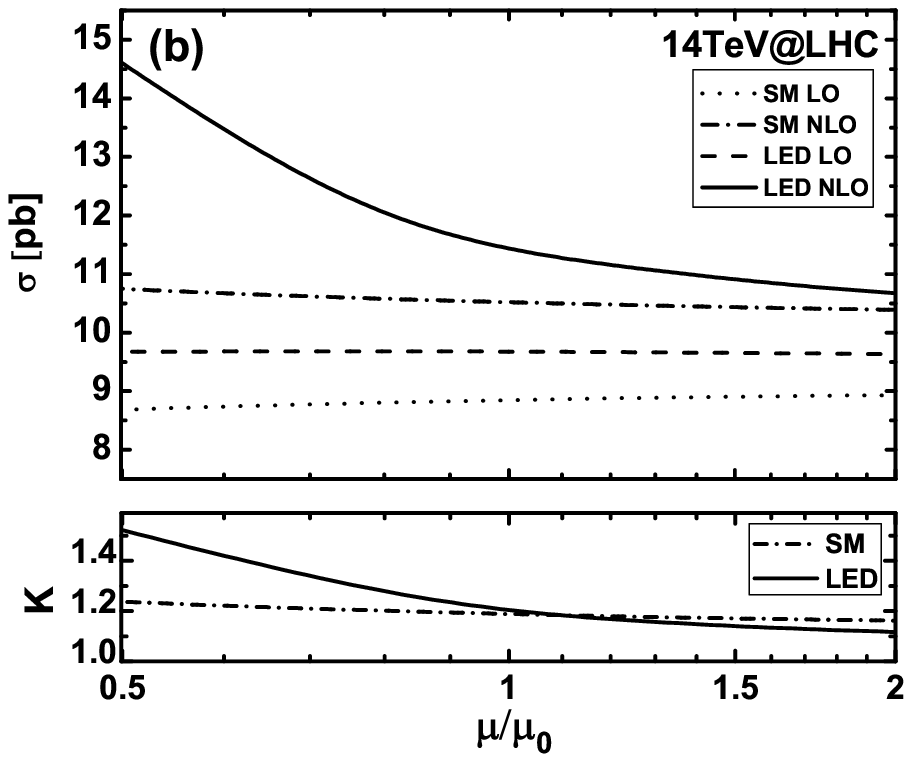}%
\hspace{0in}%
\caption{\label{fig8} The scale dependence of the LO and NLO QCD corrected
cross sections for the process \ppww in the SM and LED model, and the
corresponding $K$-factor $[K(\mu)\equiv \frac{\sigma_{NLO}(\mu)}
{\sigma_{LO}(\mu)}]$ with $M_{S}=3.5~TeV$ and $\delta =3$.
(a) At the $\sqrt{s}=7~TeV$ LHC. (b) At the $\sqrt{s}=14~TeV$ LHC. }
\end{center}
\end{figure}

In Figs.\ref{fig8}(a,b), the upper panels show the scale ($\mu$)
dependence of the LO and the NLO QCD corrected cross sections in the
SM and LED model at the $\sqrt{s}=7~GeV$ and $\sqrt{s}=14~TeV$ LHC
separately, and the corresponding $K$-factors $(K(\mu)\equiv
\frac{\sigma_{NLO}(\mu)}{\sigma_{LO}(\mu)})$ are illustrated in the
lower panels. There we take $M_{S}=3.5~TeV$ and $\delta =3$. The
scale-dependent $K$-factor in the LED model varies from $1.18$
($1.53$) to $1.19$ ($1.11$) when $\mu$ goes from $0.5\mu_0$ to
$2\mu_0$ at the $\sqrt{s}=7~TeV$ LHC (the $\sqrt{s}=14~TeV$ LHC). We
see from these upper panels that the NLO QCD corrections in the SM
and LED model do not reduce the factorization/renormalization scale
uncertainty, especially at the $\sqrt{s}=14~TeV$ LHC. That is
because the LO result underestimates the scale dependence due to the
LO contribution being from pure electroweak partonic processes. And
we find that the $K$-factors plotted in the lower figures keep the
convergence of the perturbative series in the plotted $\mu$ range at
both the $\sqrt{s}=7~TeV$ and $\sqrt{s}=14~TeV$ LHC. In further
calculations we fix $\mu = m_{W}$.

\par
\begin{figure}[htbp]
\includegraphics[scale=0.8]{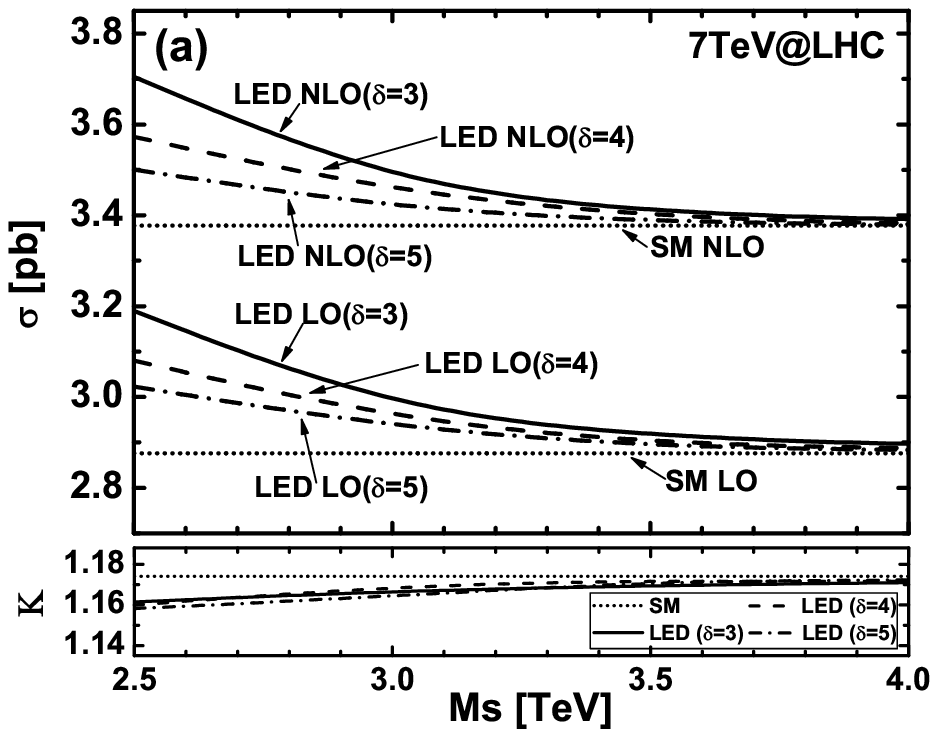}%
\hspace{0in}%
\includegraphics[scale=0.8]{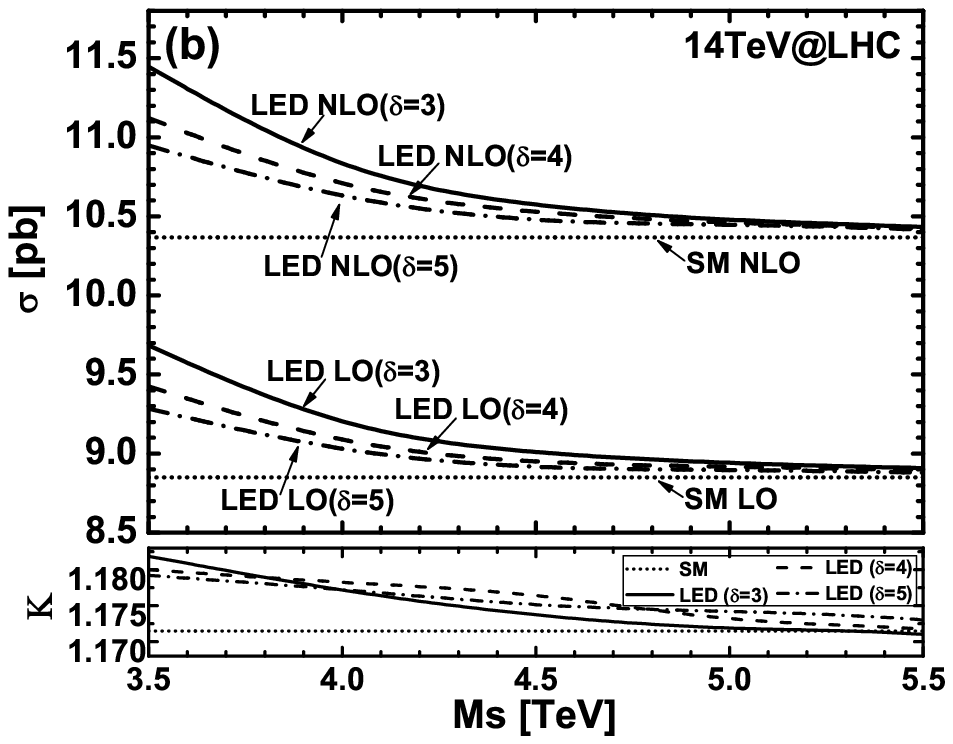}%
\hspace{0in}%
\caption{\label{fig9} The LO and NLO QCD corrected cross sections
and the corresponding $K$-factors for the process \ppww in the LED
model as functions of $M_{S}$ with $\mu=m_W$ and $\delta =3,4,5$.
(a) At the $\sqrt{s}=7~TeV$ LHC. (b) At the $\sqrt{s}=14~TeV$ LHC. }
\end{figure}

In Figs.\ref{fig9}(a,b), we depict the LO and NLO QCD corrected
cross sections and the corresponding $K$-factors for the process
\ppww in the LED model as the functions of the fundamental scale
$M_{S}$, with $\mu=m_W$ and the extra space dimension number
$\delta$ being $3$, $4$, and $5$, respectively. From the figures one
can find that the more distinct LED effect exhibits with the smaller values of $M_{S}$ and $\delta$.

\par
\begin{figure}[htbp]
\includegraphics[width=3.2in,height=3in]{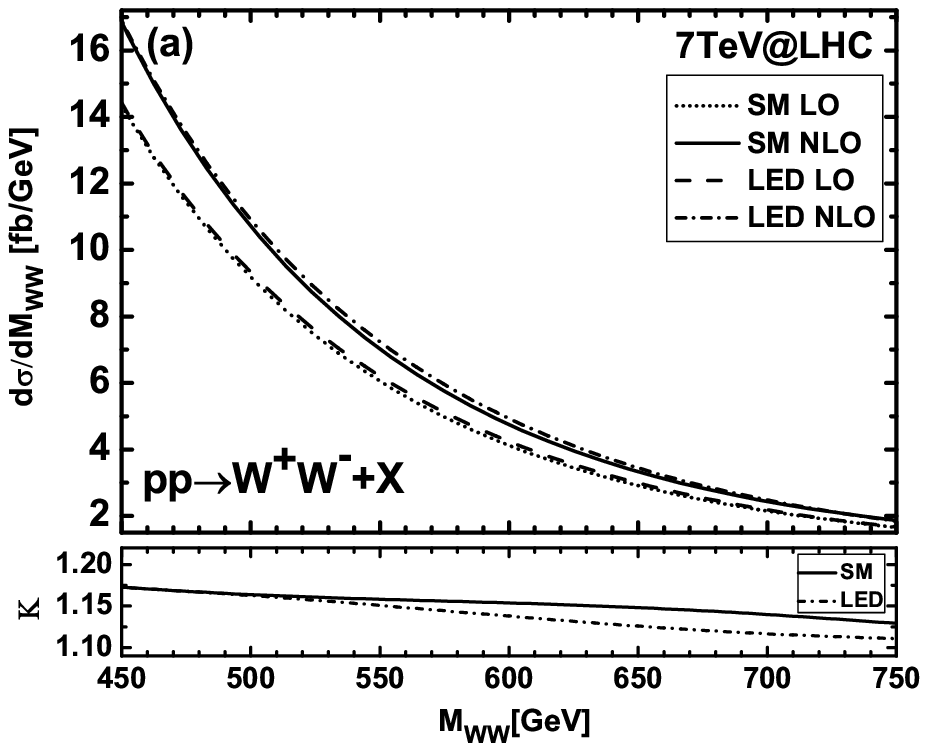}%
\hspace{0in}%
\includegraphics[width=3.2in,height=3in]{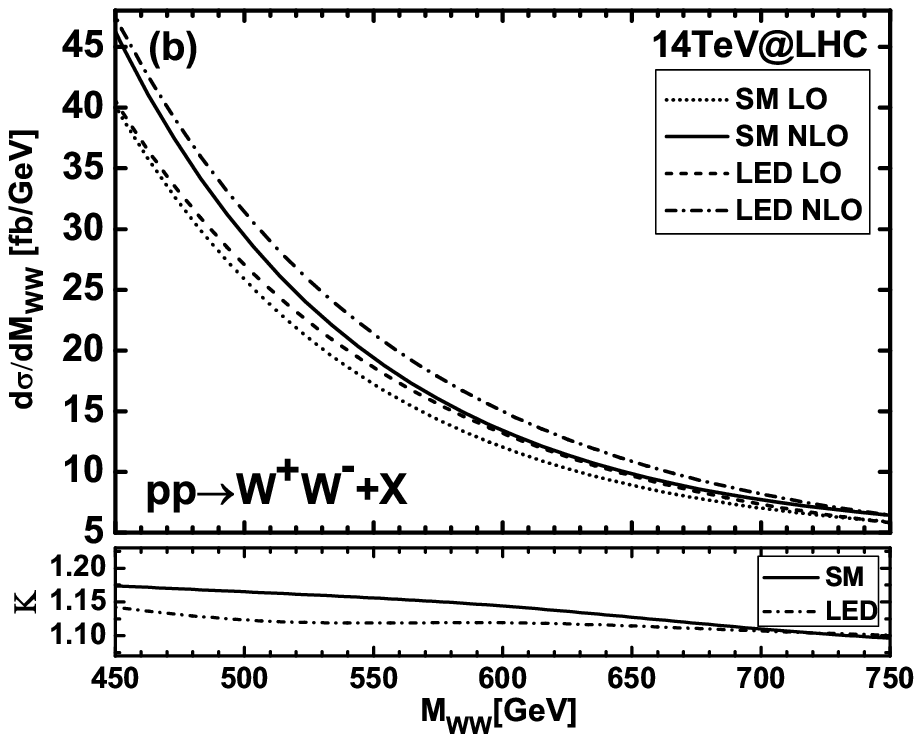}%
\hspace{0in}%
\caption{\label{fig10-1} The LO and NLO QCD corrected distributions
of the $W$-pair invariant mass $(d\sigma_{LO}/dM_{WW}$,
$d\sigma_{NLO}/dM_{WW})$ and the corresponding $K$-factors for $pp
\to W^{+}W^{-}+X$ with $M_{S}=3.5~TeV$, $\mu=m_W$ and $\delta =3$ in
the SM and LED model. (a) At the $\sqrt{s}=7~TeV$ LHC. (b) At the
$\sqrt{s}=14~TeV$ LHC. }
\end{figure}

The LO and NLO QCD corrected distributions of the $W$-pair invariant
mass and the corresponding $K$-factors ($K(M_{WW})\equiv
\frac{d\sigma_{NLO}}{dM_{WW}}/\frac{d\sigma_{LO}}{dM_{WW}}$) for the
process $pp \to W^{+}W^{-}+X$ in the SM and LED model at the
early and future LHC, are shown in Figs.\ref{fig10-1}(a) and (b),
separately. There the results are for $M_{S}=3.5~TeV$, $\mu=m_W$, at
a fixed value 3 for the number of extra dimensions and obtained by
taking the input parameters and the event selection constraints mentioned
above. As we expected, the LO and NLO QCD corrected differential
cross sections of the $W$-pair invariant mass become less with the
increment of $M_{WW}$.

\par
\begin{figure}[htbp]
\includegraphics[width=3.2in,height=3in]{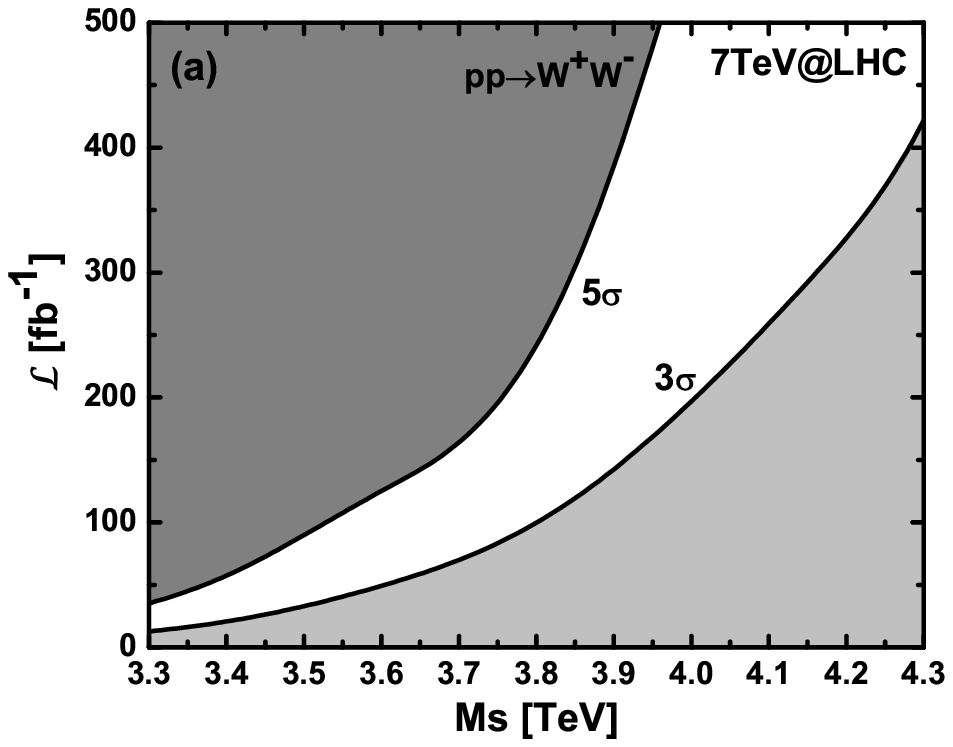}%
\hspace{0in}%
\includegraphics[width=3.2in,height=3in]{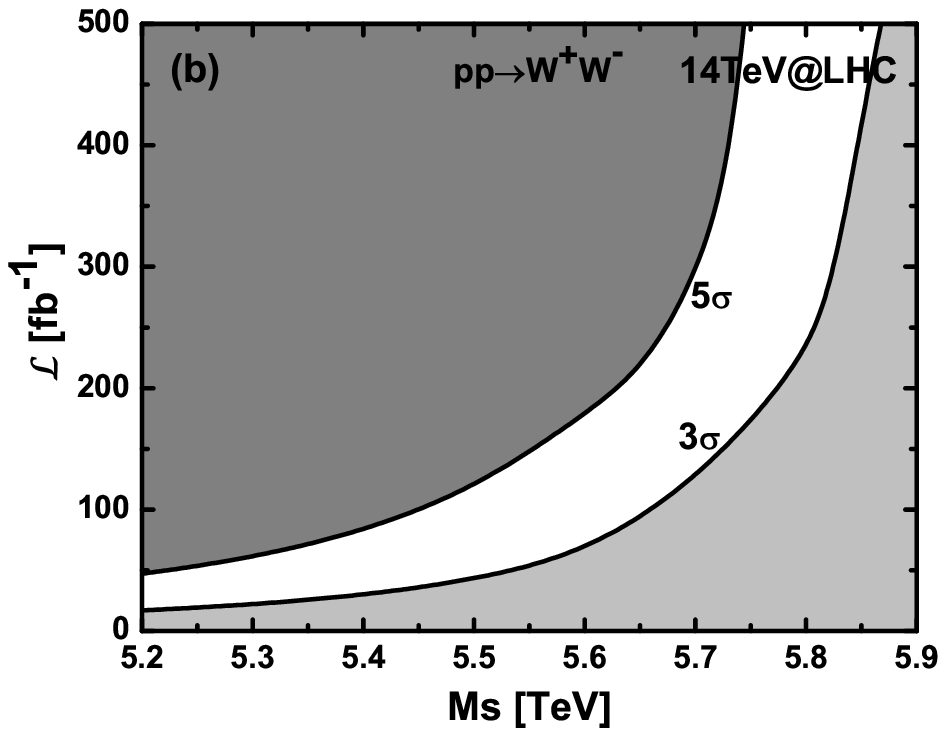}%
\hspace{0in}%
\caption{\label{fig10} The LED effect discovery area (dark) and the
exclusion area (gray) in the $\mathcal{L}-M_{S}$ space for the \ppww
process. (a) At the $\sqrt{s}=7~TeV$ LHC. (b) At the
$\sqrt{s}=14~TeV$ LHC.  }
\end{figure}

It is clear that if the deviation of the cross section from the SM
prediction is large enough, the LED effect including the NLO QCD
corrections can be found. We assume that the LED effect can and
cannot be observed, only if
\begin{eqnarray}
\label{upper} \Delta\sigma_{NLO}
=|\sigma_{NLO}^{LED}-\sigma_{NLO}^{SM}| \geq
\frac{5\sqrt{\mathcal{L}\sigma_{NLO}^{LED}}}{\mathcal{L}}\equiv
5\sigma
\end{eqnarray}
and
\begin{eqnarray}
\label{lower} \Delta\sigma_{NLO}
=|\sigma_{NLO}^{LED}-\sigma_{NLO}^{SM}| \leq
\frac{3\sqrt{\mathcal{L}\sigma_{NLO}^{LED}}}{\mathcal{L}}\equiv
3\sigma,
\end{eqnarray}
respectively. In Figs.\ref{fig10}(a,b), we present the discovery and
exclusion regions in the luminosity-fundamental scale space
($\mathcal{L}-M_{S}$) for the \ppww process with $\delta = 3$.
Figure.\ref{fig10}(a) is for the $\sqrt{s}=7~TeV$ LHC and
Fig.\ref{fig10}(b) for the $\sqrt{s}=14~TeV$ LHC, where the LED
effect can and cannot be observed in the dark and gray- region,
separately. We list some typical data which are read out from
Figs.\ref{fig10}(a,b) in Table \ref{tab3}. There the discovery and
exclusion fundamental scale $M_S$ values at the early and future LHC
are presented. It shows that by using the $W$-boson pair production
events we could set an exclusion limit on the cutoff scale $M_S$ to be
$1.80~TeV$ at the $95\%$ confidence level at the $\sqrt{s} = 7~TeV$
LHC with an integrated luminosity of $36~(pb)^{-1}$. This is in the
$M_S$ lower limit range of $ 1.6\sim 2.3~TeV$ obtained
experimentally by the CMS using the diphoton final state data
samples \cite{CMS-1}.
\begin{table}[htbp]
\begin{center}
\begin{tabular}{|c|c|c|c|c|}
\hline
{}Luminosity$(\mathcal{L})$              &\multicolumn{2}{c|}{$\sqrt{s}=7~TeV$}    &\multicolumn{2}{c|}{$\sqrt{s}=14~TeV$}       \\ \cline{2-5}
{}($\mathcal{L}$)         &$M_{S}[TeV]$($3\sigma$)&$M_{S}[TeV]$($5\sigma$)  &$M_{S}[TeV]$($3\sigma$)&$M_{S}[TeV]$($5\sigma$) \\
\hline  100 ${fb}^{-1}$   &3.83                   &3.50                     &5.69                   &5.46                 \\
\hline  200 ${fb}^{-1}$   &3.98                   &3.74                     &5.79                   &5.62                 \\
\hline  300 ${fb}^{-1}$   &4.17                   &3.85                     &5.83                   &5.70                 \\
\hline  36  ${pb}^{-1}$   &1.80                   &1.68                     &2.03                   &1.89                 \\
\hline
\end{tabular}
\caption{  \label{tab3} The discovery ($\Delta\sigma_{tot} \geq
5\sigma$) and exclusion ($\Delta\sigma_{tot} \leq 3\sigma$) LED
model fundamental scale ($M_S$) values for the \ppww process at the
early ($\sqrt{s}=7~TeV$) and future ($\sqrt{s}=14~TeV$) LHC. }
\end{center}
\end{table}

\par
\begin{figure}[htbp]
\includegraphics[width=3.2in,height=3in]{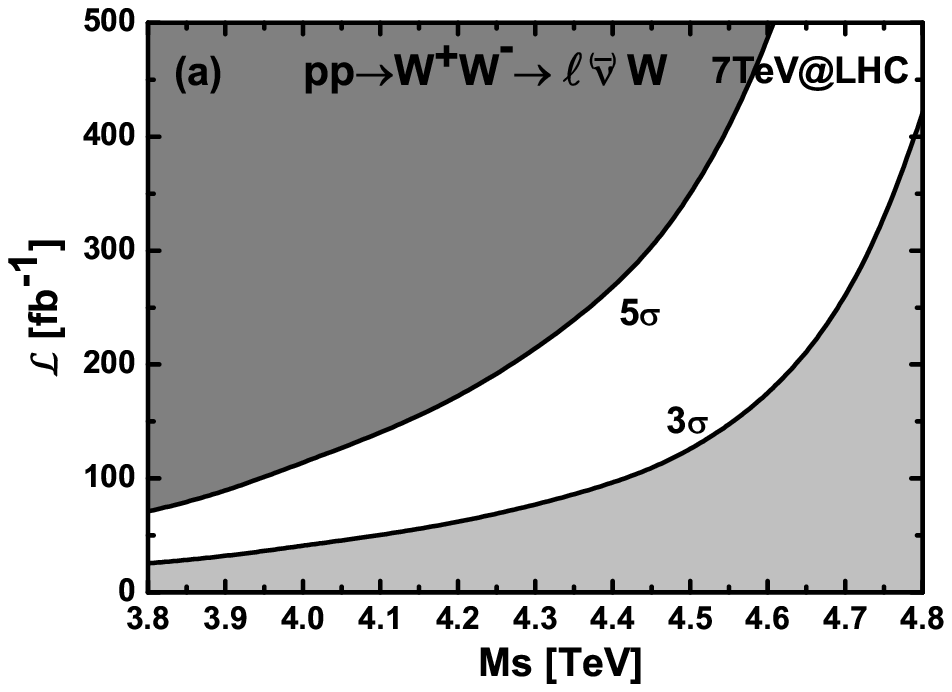}%
\hspace{0in}%
\includegraphics[width=3.2in,height=3in]{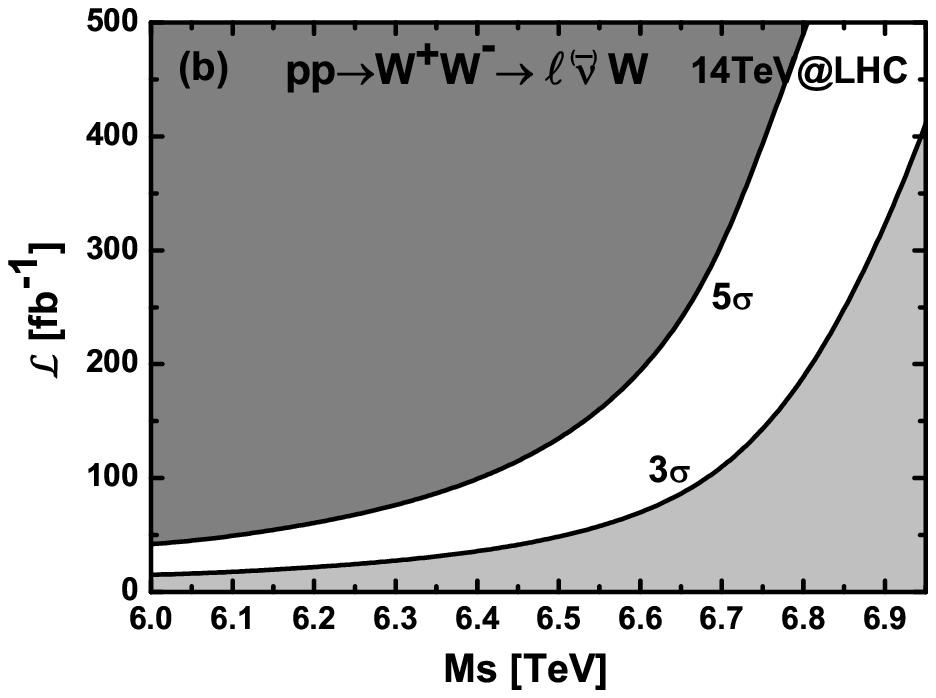}%
\hspace{0in}%
\caption{\label{fig11} The LED effect discovery area (dark) and
exclusion area (gray) in the $\mathcal{L}-M_{S}$ space for the
\ppwlv ($\ell = e, \mu$) processes with the constraints of
$M_{WW}>400~GeV$, $p_{T}^{l} > p_{T,l}^{cut}=100~GeV$, and the jet
event selection criterion declared above. (a) at the
$\sqrt{s}=7~TeV$ LHC. (b) at the $\sqrt{s}=14~TeV$ LHC. }
\end{figure}

Now we consider the subsequential leptonic (electron, muon) decay of
one of the two $W$-bosons. In collecting the \ppwlv ($\ell = e,
\mu$) events we do not distinguish the leptonic charge. We fix the
branching fraction for $W$-boson decay ($W^{\mp} \to \ell^{\mp}
\stackrel{(-)}{\nu},~\ell = e, \mu$) as $21.32\%$ \cite{18},
$\mathcal{L}=300~fb^{-1}$, and take the number of the extra
dimensions $\delta = 3$, the constraints of $M_{WW}>400~GeV$,
$p_{T}^{l} > p_{T,l}^{cut}=100~GeV$, and the jet event selection
criterion as declared above. We show the discovery and exclusion
regions in the $\mathcal{L}-M_{S}$ space for the processes \ppwlv
($\ell = e, \mu$) in Figs.\ref{fig11}(a) and (b) for the
$\sqrt{s}=7~TeV$ and $\sqrt{s}=14~TeV$ LHC, respectively, The dark
and gray- regions represent the parameter space where the LED effect
can and cannot be observed separately. Some representative data for
the discovery and exclusive fundamental scale $M_S$ values at the
early ($\sqrt{s}=7~TeV$) and future ($\sqrt{s}=14~TeV$) LHC read out
from Figs.\ref{fig11}(a,b) are presented in Table \ref{tab4}.
We can see from the table that by using \ppwlv ($\ell = e, \mu$)
processes with the constraints of $M_{WW}>400~GeV$, $p_{T}^{l} >
p_{T,l}^{cut}=100~GeV$, and our chosen jet event selection
criterion, we could get exclusion lower limit on $M_S$ as $2.19~TeV$
at the $95\%$ confidence level at the $\sqrt{s} = 7~TeV$ LHC with an
integrated luminosity of $36~(pb)^{-1}$, which is larger than that
obtained by analyzing the $W$-pair production events as described above.
\begin{table}[htbp]
\begin{center}
\begin{tabular}{|c|c|c|c|c|}
\hline {}Luminosity$(\mathcal{L})$              &\multicolumn{2}{c|}{$\sqrt{s}=7TeV$}
&\multicolumn{2}{c|}{$\sqrt{s}=14TeV$}       \\ \cline{2-5}
{}($\mathcal{L}$)         &$M_{S}[TeV]$($3\sigma$)&$M_{S}[TeV]$($5\sigma$)  &$M_{S}[TeV]$($3\sigma$)&$M_{S}[TeV]$($5\sigma$) \\
\hline  100 ${fb}^{-1}$   &4.42                 &3.95                &6.69                 &6.41                \\
\hline  200 ${fb}^{-1}$   &4.65                 &4.29                &6.82                 &6.62                 \\
\hline  300 ${fb}^{-1}$   &4.74                 &4.45                &6.87                 &6.71                 \\
\hline  36  ${pb}^{-1}$   &2.19                 &1.96                 &2.98                 &2.80                 \\
\hline
\end{tabular}
\caption{  \label{tab4} The discovery ($\Delta\sigma_{tot} \geq
5\sigma$) and exclusion ($\Delta\sigma_{tot} \leq 3\sigma$)
fundamental scale ($M_{S}$) values for the \ppwlv ($\ell = e, \mu$)
processes in the $\mathcal{L}-M_{S}$ space for the \ppwlv ($\ell =
e, \mu$) processes with the constraints of $M_{WW}>400~GeV$ and
$p_{T}^{l} > p_{T,l}^{cut}=100~GeV$: at the $\sqrt{s}=7~TeV$
LHC and at the $\sqrt{s}=14~TeV$ LHC. }
\end{center}
\end{table}

\par
\begin{figure}[htbp]
\includegraphics[width=3.2in,height=3in]{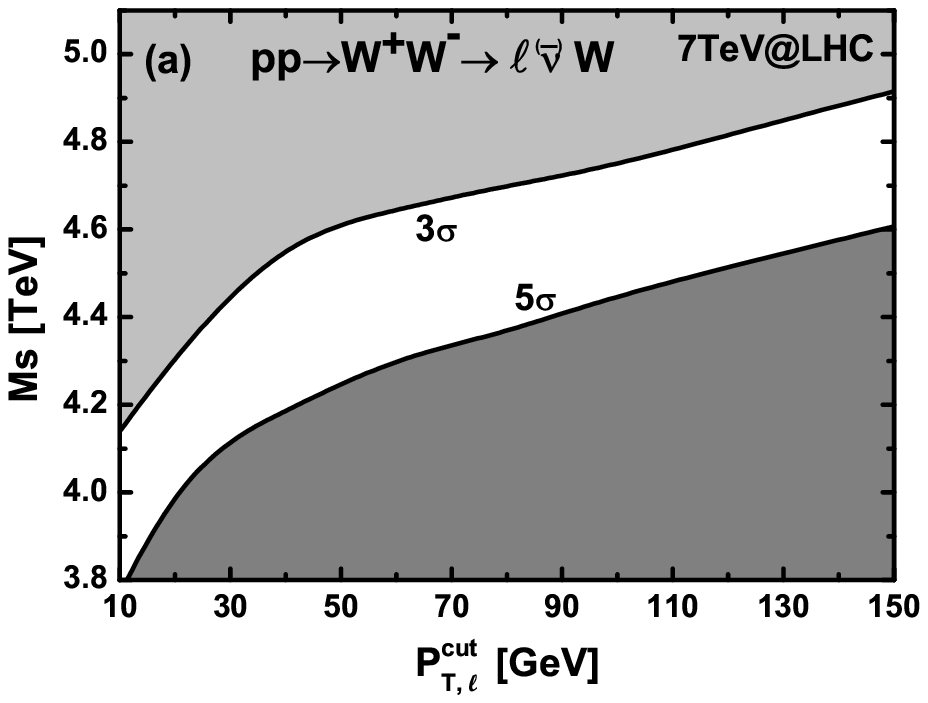}%
\hspace{0in}%
\includegraphics[width=3.2in,height=3in]{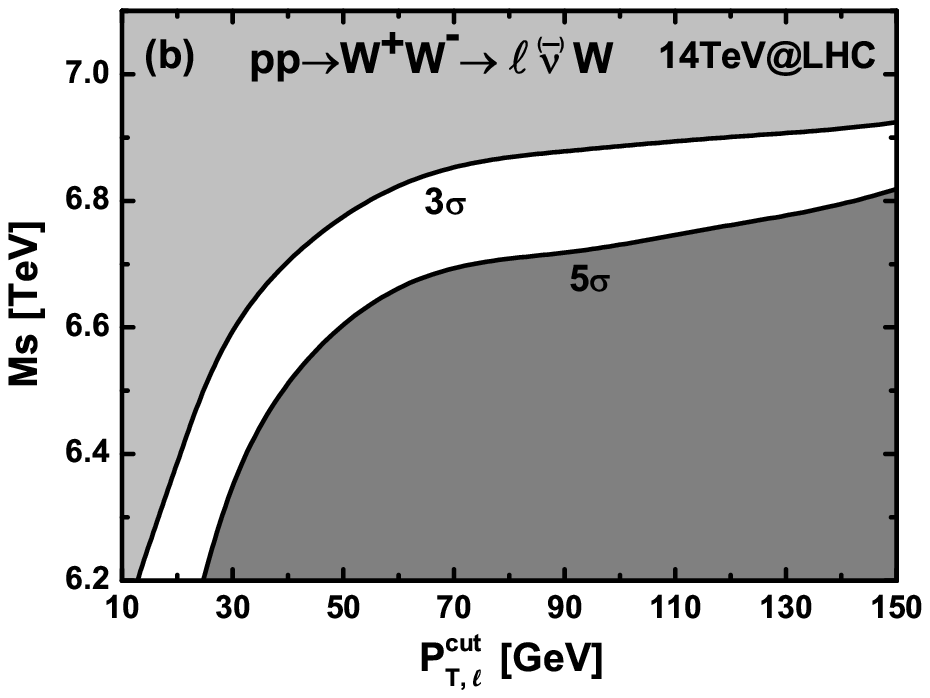}%
\hspace{0in}%
\caption{\label{fig12}  The LED effect discovery area (dark) and
exclusion area (gray) in the $M_{S}-p_{T,l}^{cut}$ space for the
\ppwlv ($\ell = e, \mu$) processes with $\delta = 3$ and
$\mathcal{L}=300~fb^{-1}$. (a) At the $\sqrt{s}=7~TeV$ LHC. (b) At
the $\sqrt{s}=14~TeV$ LHC. }
\end{figure}

We depict the LED discovery and exclusion regions in the
$M_{S}-p_{T,l}^{cut}$ space for the processes \ppwlv ($\ell = e,
\mu$) in Figs.\ref{fig12}(a) and (b) with $\delta = 3$,
$\mathcal{L}=300~fb^{-1}$, $M_{WW}>400~GeV$, and the branching
fraction for $W$-boson decays ($W^{\mp} \to \ell^{\mp}
\stackrel{(-)}{\nu},~\ell = e, \mu$) as $21.32\%$, where
Fig.\ref{fig12}(a) and Fig.\ref{fig12}(b) are for the
$\sqrt{s}=7~TeV$ and $\sqrt{s}=14~TeV$ LHC respectively. The dark
and gray- regions represent the parameter regions where the LED effect
can and cannot be observed, separately, with the constraints of
$p_{T,l} > p_{T,l}^{cut}$ and the $W$-pair invariant mass $M_{WW}>
400~GeV$. Some  representative data are listed in Table \ref{tab5} for the discovery and exclusion
fundamental scale $M_S$ values with different $p_{T,l}^{cut}$ values
at the $\sqrt{s}=7~TeV$ and $\sqrt{s}=14~TeV$ LHC as shown in
Figs.\ref{fig12}(a,b). We can see
that in the case where we fix the integral luminosity (e.g.
$\mathcal{L}=300~fb^{-1}$), we could improve slightly the low limit
on $M_S$ if we adopt a larger lower cut on lepton transverse momentum
($p_{T,l}^{cut}$).
\begin{table}[htbp]
\begin{center}
\begin{tabular}{|c|c|c|c|c|}
\hline
{}$p_T^l$ cut value      &\multicolumn{2}{c|}{$7~TeV$}    &\multicolumn{2}{c|}{$14~TeV$}       \\ \cline{2-5}
{}($p_{T,l}^{cut}$)         &$M_{S}[TeV]$($3\sigma$)&$M_{S}[TeV]$($5\sigma$)  &$M_{S}[TeV]$($3\sigma$)&$M_{S}[TeV]$($5\sigma$) \\
\hline  50 GeV   &4.61                 &4.24                &6.67                 &6.60                \\
\hline  100 GeV  &4.74                 &4.45                &6.87                 &6.71                 \\
\hline  150 GeV  &4.92                 &4.61                &6.92                 &6.80                 \\
\hline
\end{tabular}
\caption{  \label{tab5} The discovery ($\Delta\sigma_{tot} \geq
5\sigma$) and exclusion ($\Delta\sigma_{tot} \leq 3\sigma$) LED
model fundamental scale ($M_{S}$) values in the
$M_{S}-p_{T,l}^{cut}$ space for the \ppwlv ($\ell = e, \mu$)
processes with the constraints of $M_{WW}>400~GeV$ and $p_{T}^{l} >
p_{T,l}^{cut}$. (a) at the $\sqrt{s}=7~TeV$ LHC. (b) at the
$\sqrt{s}=14~TeV$ LHC. }
\end{center}
\end{table}

\vskip 5mm
\section{Summary}
\par
We calculate the NLO QCD corrections to the $pp \to W^+W^- \to
W^{\pm}l^{\mp}\stackrel{(-)}{\nu} + X$ process in the SM and LED
model at the LHC. We investigate the integrated cross sections, the
distributions of some kinematic variables and how they are affected
by radiative corrections. The calculations are compared with
previous works, and finally the reliable numerical results are
obtained. We find that the NLO QCD corrected results do not show
remarkable reduction of the scale uncertainties of the LO cross
sections in both the SM and LED model, because the uncertainty of
the LO cross section is underestimated. The scale-dependent
$K$-factor is found to be the value from $1.18$ ($1.53$) to $1.19$
($1.11$) when $\mu$ goes from $0.5\mu_0$ to $2\mu_0$ at the
$\sqrt{s}=7~TeV$ ($\sqrt{s}=14~TeV$) LHC, with the constraints of
$M_{WW}>400~GeV$ and our jet event selection criterion. The
$5\sigma$ discovery and $3\sigma$ exclusion ranges for the LED
parameters $M_{S}$ are also obtained in the NLO QCD. The inclusion
of the effects of the virtual KK graviton turns out to enhance the
differential distributions of kinematical observables generally. We
conclude that the NLO QCD correction to the $W$-pair production
make it possible to precisely test the $TeV$ quantum gravity in
the LED scenario at the LHC.

\vskip 5mm
\par
\noindent{\large\bf Acknowledgments:} This work was supported in
part by the National Natural Science Foundation of China
(No.10875112, No.11075150, No.11005101), and the Specialized
Research Fund for the Doctoral Program of Higher
Education(No.20093402110030).

\end{document}